\documentclass[aps,pre,twocolumn,showpacs,amsmath]{revtex4}
\usepackage{graphicx} \usepackage{verbatim}
\usepackage[pdftex]{hyperref}
\hypersetup{colorlinks=true,linkcolor=blue,citecolor=blue,urlcolor=blue}

\begin{document}
\title{Transition state theory and the dynamics of hard disks}
\author{M. Barnett-Jones, P. A. Dickinson, M. J. Godfrey, T. Grundy  and M. A. Moore}
\affiliation{School of Physics and Astronomy, University of Manchester,
 Manchester M13 9PL, UK}

\begin{abstract}
The dynamics  of two and  five disk systems  confined in a  square has
been studied  using molecular  dynamics simulations and  compared with
the  predictions  of  transition   state  theory.   We  determine  the
partition functions $Z$ and  $Z^{\ddagger}$ of transition state theory
using a  procedure first used by  Salsburg and Wood  for the pressure.
Our simulations show this procedure and transition state theory are in
excellent  agreement with  the simulations.   A generalization  of the
transition state theory to the case  of a large number of disks $N$ is
made  and shown  to be  in full  agreement with  simulations  of disks
moving in  a narrow channel.  The  same procedure for  hard spheres in
three  dimensions  leads to  the  Vogel--Fulcher--Tammann formula  for
their alpha relaxation time.

\end{abstract}
\maketitle


\section{Introduction}

The long relaxation times seen in supercooled liquids have long been a
challenge  to understand  \cite{AG,Biroli2011}.   Glassy behavior  has
been extensively modelled by studying hard spheres in three dimensions
and hard  disks in two  dimensions. Some of  this work is  reviewed in
Ref. \cite{ZamponiParisi}.   In this paper  we study small  systems of
disks, in  particular two disks and  five disks confined  in a square,
first  using event  driven molecular  dynamics  and then  by means  of
transition  state theory  \cite{BowlesMonPercus,Orland}. In  the final
section of  the paper we use  the insights gained  from studying small
systems  to speculate  about the  behavior  of large  numbers of  hard
spheres or disks.

It  is   convenient  from  the  outset  to   introduce  the  following
terminology.  The transition state  is the neck in configuration space
through which the  system has to pass to escape  its initial state.  A
\emph{configuration}  of  the  $N$  disks  is defined  by  the  $N  d$
coordinates   of  the   disk   centers  (for   disks,  $d=2$);   every
configuration  belongs  to  a   \emph{state},  which  is  the  set  of
configurations  that can  be  reached from  it  without violating  the
no-overlap  constraint appropriate  for hard  disks and  spheres.  The
transition state  theory will be found  to work well when  the neck is
narrow, that is, when there are long relaxation times in the system.

We  shall   illustrate  the  process   of  escape  from   the  initial
configuration for two simple systems, consisting of either two or five
disks  confined in a  square.  For  these simple  systems we  can make
explicit  the narrow necks  in configuration  space through  which the
system  can escape  from its  initial configuration  near  an inherent
state~\cite{StillingerDiMarzioKornegay}.     We   shall    show   that
transition state  theory provides a  quantitative account of  the slow
relaxational processes  in these  small systems.  The  theory requires
that one evaluates  a variant of the partition  function of the system
at the  neck ($Z^{\ddagger}$)  and to do  this we adopt  the procedure
first used by Salsburg and Wood \cite{SW} to calculate the pressure of
hard spheres near their largest  packing density.  We have checked its
accuracy for these small systems  by comparing its predictions for the
pressure of the system and  the relaxation times with results obtained
directly from event driven molecular dynamics.

Of course, one is only interested in small systems of disks because of
the light  their study might shine  on large systems of  hard disks or
spheres. We  shall show that as  $N$, the number of  disks or spheres,
becomes large, then, under certain circumstances, our transition state
formula  for  the relaxation  time  in the  system  goes  over to  the
well-known    Vogel--Fulcher--Tammann     (VFT)    equation.     These
circumstances are  evidently realized  in at least  one case,  that of
disks  moving in  a narrow  channel \cite{Ivan,  Mahdi,  Bowles}.  The
agreement is  quantitative in this case  \cite{GodfreyMoore}. For hard
spheres in three dimensions their  alpha relaxation time can be fitted
by the VFT form \cite{Brambilla}, but with the divergence occurring at
a  density  below  that  of  random close  packing.   This  matter  is
discussed in Sec.~\ref{extensionlargeN}, where we then go on to give a
speculative extension  to our procedure  which leads to  a generalized
VFT formula where  the divergence takes place at  a density similar to
that of random close packing.

Throughout our study  of two and five disks confined  in a square, $L$
denotes  the length  of one  side of  the square  and $r$  denotes the
radius of the  disks.  The packing fraction $\phi$  is then defined as
the fraction of  the area of the square that is  covered by the disks,
$\phi  =   N\pi  r^2/L^2$.   The   two  disk  system  is   studied  in
Sec.~\ref{sec:2disk}{}.  In Sec.~\ref{5diskconfig} the inherent states
and the necks in configuration space which separate them are discussed
for the  five disk system. In Sec.~\ref{5diskdynamics}  we compare the
results of  our event driven  molecular dynamics simulations  for five
disks with the predictions of transition state theory.

\section{The two-disk system}
\label{sec:2disk}
\begin{figure}
\begin{center}
\includegraphics[width=3.5in]{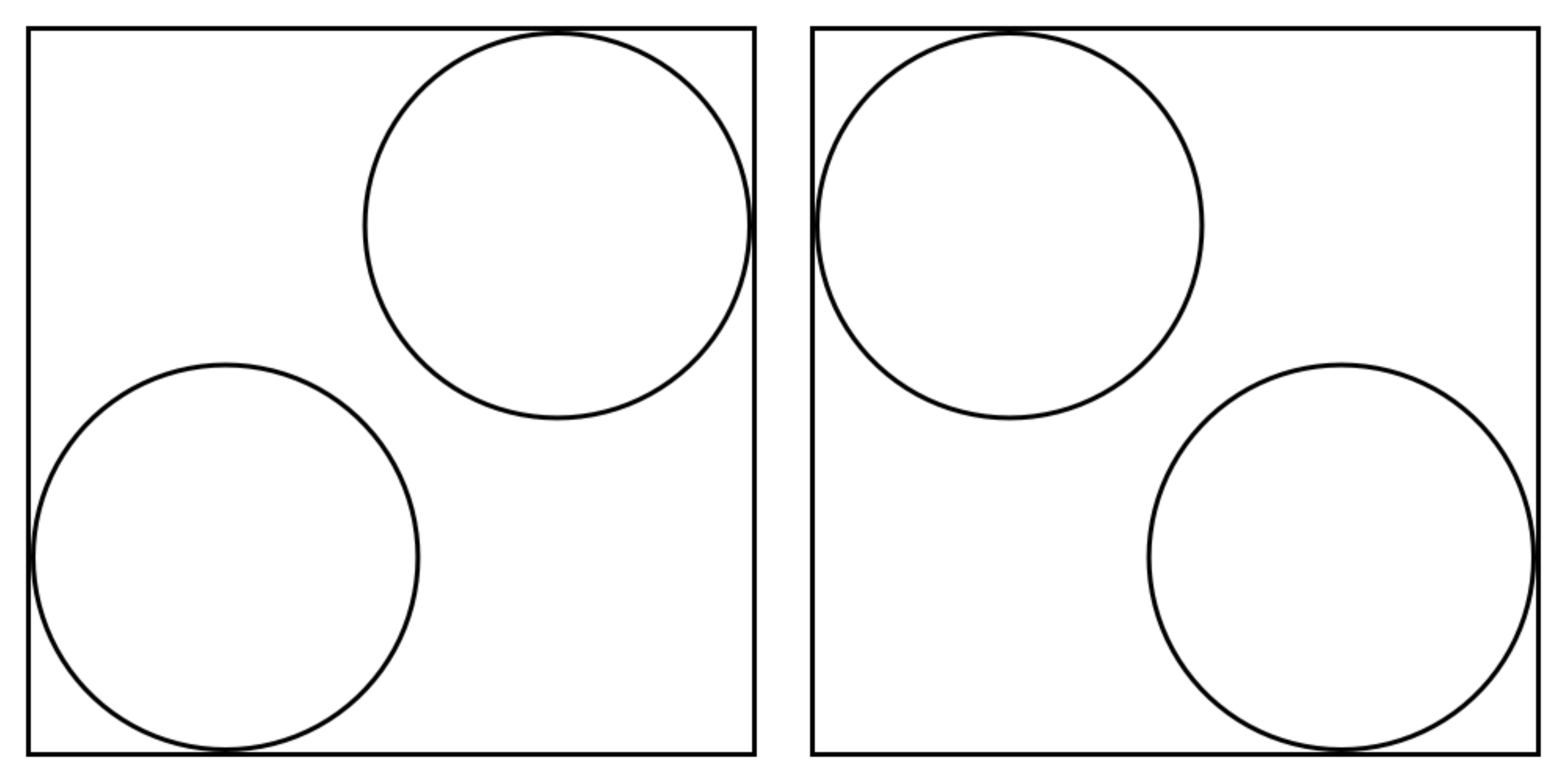}
\caption{Mutually  inaccessible  configurations  of  a system  of  two
  disks,  illustrated  for  $r=0.27\,L$,  which is  greater  than  the
  critical  value $r=0.25\,L$  below  which the  disks  can pass  each
  other.}
\label{twodisks}
\end{center}
\end{figure}

\begin{figure}
\begin{center}
\includegraphics[width = 3.5in]{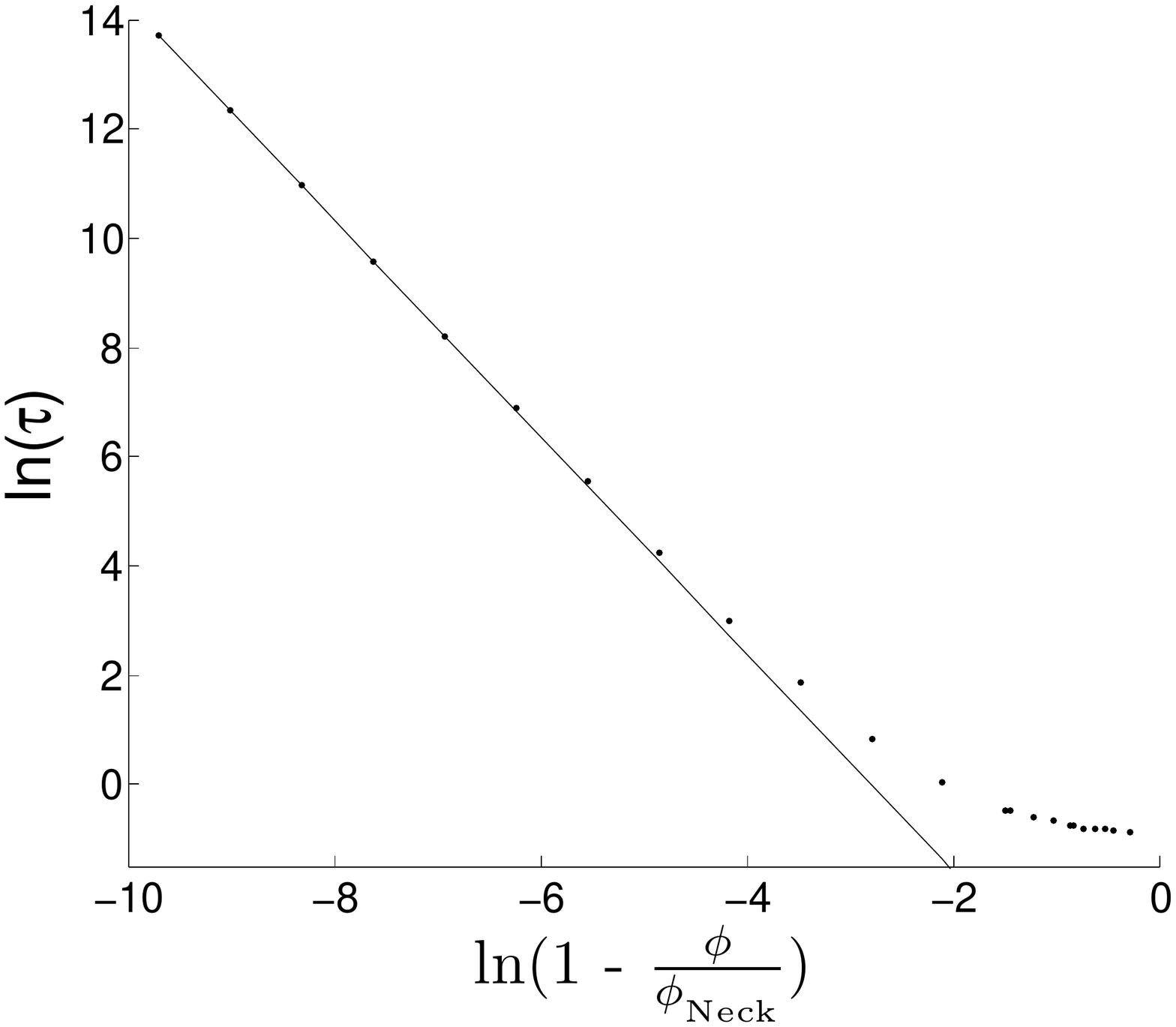}
\caption{Molecular   dynamics  transition  rates   ($1/\tau$)  between
  metastable ``glassy'' states of  the two-disk system, as the packing
  fraction,     $\phi$,      approaches     the     critical     value
  $\phi_{\text{Neck}}\simeq0.3927$ corresponding  to $r=0.25\,L$.  The
  gradient  of  the straight-line  fit  to  the  last five  points  is
  $-1.986$, which is  close to the value $-2$  predicted by transition
  state theory.}
\label{lograte}
\end{center}
\end{figure}
Two  disks confined  to a  square have  been considered  previously by
Speedy \cite{Speedy}.  For $r<L/4$ (or $\phi< \phi_{\text{Neck}}\simeq
0.3927$)  the disks  can pass  each  other, though  this becomes  more
difficult  as  $r\rightarrow  L/4$ (see  Fig.~\ref{twodisks}).   Awazu
\cite{Awazu}  studied an  autocorrelation function  which  developed a
plateau in  this limit, and  which he argued showed  similarities with
the $\alpha$  and $\beta$ relaxation  processes found in  glasses. For
larger  values  of  $r$,  the configuration  space  (disregarding  the
identity of the  disks) is broken into two  states.  For $r\rightarrow
L/(2+\sqrt2\,)$,  or $\phi\rightarrow\phi_J\simeq0.5390$,  the maximum
density possible,  the disk  centers lie on  the same diagonal  of the
square.   These (two) limiting  configurations are  the \emph{inherent
  structures} introduced in Ref.~\cite{StillingerDiMarzioKornegay}.

Speedy \cite{Speedy} has considered  the thermodynamics of this system
and,  in particular,  finds weakly  non-analytic contributions  to the
thermodynamic  quantities, such  as the  pressure, at  $r=L/4$. Speedy
used transition state theory to determine the alpha relaxation time of
the  system, which  according to  Awazu, is  the time  $\tau$  to flip
between the  two configurations.  The origin of  this behavior  can be
obtained  from transition state  theory \cite{BowlesMonPercus,Orland}.
In  this  well-studied  approximation,   which  works  best  when  the
transition  rate over  a barrier  is  small, the  transition rate  $R$
between two states varies as
\begin{equation}
R = 1/\tau  \sim v \,\frac{Z^{\ddagger}}{Z},
\label{transitionrate}
\end{equation}
where  $v$ is  a  typical  particle speed  and  $Z^{\ddagger}$ is  the
partition  function evaluated  at the  top  of the  barrier along  the
trajectory which  separates the  states; see Ref.~\cite{Orland}  for a
full description of the  transition state formalism and the definition
of $Z^{\ddagger}$.  In  the case of two disks  passing this means that
instead of  the full partition function integral  over $(x_1,y_1)$ and
$(x_2,y_2)$, there  is a constraint  that $y_1=y_2=y$, (say) so  it is
effectively a three-dimensional integral.  The integral over $y$ gives
a  trivial factor  $(L-2r)$ and  the  remaining two  integrals give  a
factor   $(1-\phi/\phi_{\text{Neck}})^2$  in   the  limit   $\phi  \to
\phi_{\text{Neck}}$  by  the  argument   used  by  Salsburg  and  Wood
\cite{SW}.  $Z$~itself is  essentially just a constant: it  has a very
mild    singularity,    $Z_{\text{reg}}+C(1   -    \phi_{\text{Neck}}/
\phi)^{5/2}$,   when   the    packing   fraction   $\phi$   approaches
$\phi_{\text{Neck}}$   from  above~\cite{Speedy}.    Transition  state
theory thus  predicts a  slope of $-2$  in Fig.~\ref{lograte}  for the
dependence  of the  relaxation time  on packing  fraction  as $\phi\to
\phi_{\text{Neck}}$.  (The  full integrals for  $Z^{\ddagger}$ and $Z$
were explicitly evaluated  by Speedy \cite{Speedy}.)  Our event-driven
molecular  dynamics results  (Fig.~\ref{lograte}) are  consistent with
the   transition    state   theory   prediction    that   $\tau   \sim
1/(1-\phi/\phi_{\text{Neck}})^{\alpha}$, with $\alpha=2$, for the case
of    two    disks    in     a    square    box,    in    the    limit
$\phi\to\phi_{\text{Neck}}$.

\section{Configurations of the five-disk system.}
\label{5diskconfig}
\begin{figure}
\begin{center}
\includegraphics[width=3in]{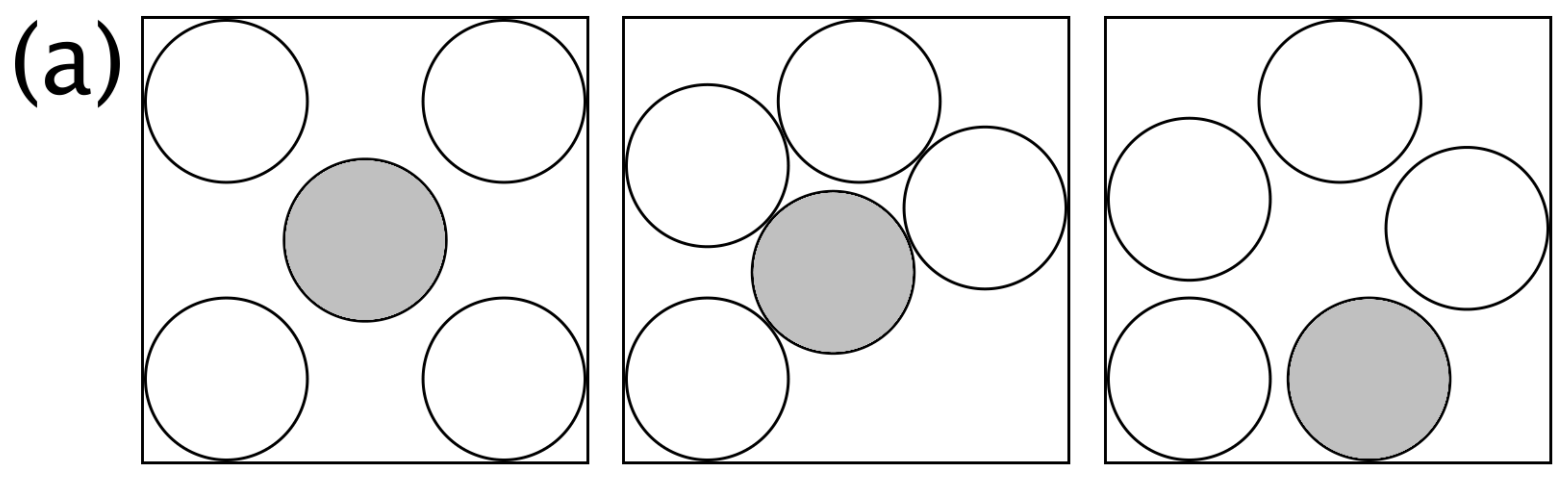}
\includegraphics[width=3in]{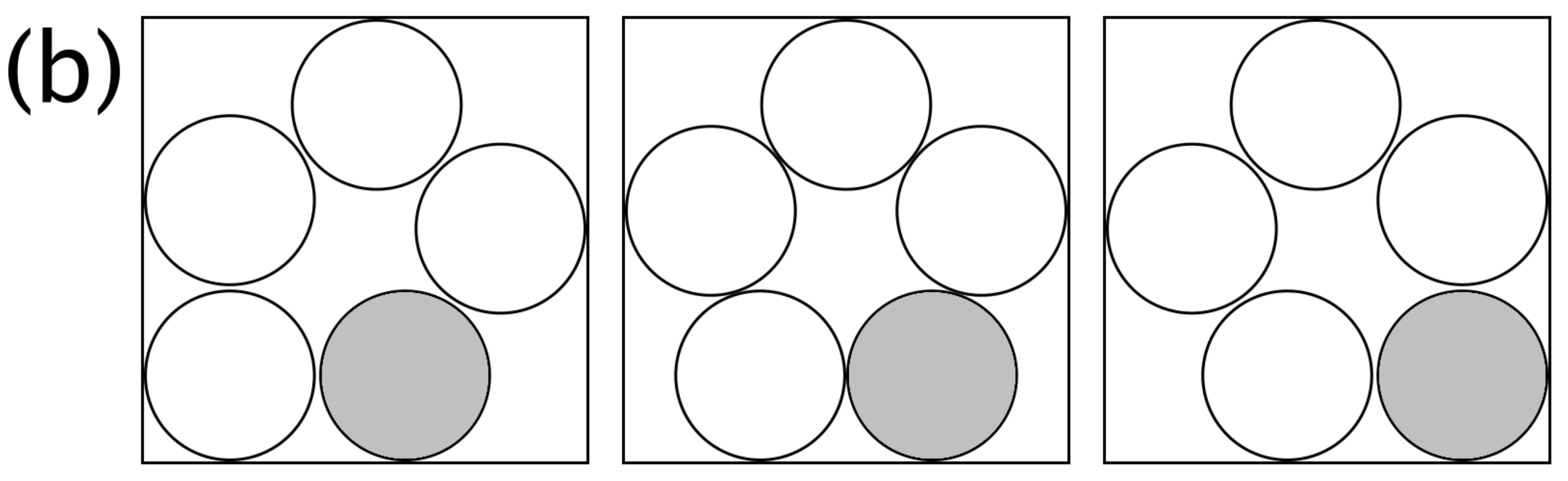}
\caption{(a)  Representative  configurations  of  the  ``crystalline''
  (left)  and  ``glassy''  (right)   states  of  a  five-disk  system,
  connected  by  a transition  state  (center),  shown  here with  the
  largest radius, $r\simeq0.1863\,L$, for which the transition between
  crystalline  and glassy states  is possible;  (b)  two frozen  glass
  states (left and right) and the transition state (center) connecting
  them, shown with the maximum radius, $r\simeq0.1942\,L$, for which a
  transition between glass states is possible.}
\label{critical}
\end{center}
\end{figure}

The configuration  space of five disks  confined to a  square has been
analyzed  previously:  Bowles   and  Speedy  \cite{SpeedyBowles}  have
discussed  the  thermodynamics and  dynamics;  Hinow \cite{Hinow}  has
studied   the  jammed  states   of  this   system;  and   Carlsson  et
al.\  \cite{Carlsson}  have  given  a  detailed analysis  of  how  the
topology  of the  configuration  space depends  on~$r$.   We refer  to
Fig.~\ref{critical} for  configurations of  the disks at  two critical
values of the radius.

Below   the    fluid--crystal   critical   point,    i.e.,   for   $r<
r_{\text{cg}}\simeq 0.1863\,L$, the system is fluid (any pair of disks
can exchange  position), but  for slightly greater  values of  $r$ the
configuration space is fractured  into two states: a ``crystal'' state
in which one disk is surrounded  by the four others, confined near the
corners of the box; and a  ``glass'' state in which all five disks lie
close to  the walls of the box  and are unable to  change their order.
Above  $r=r_{\text{gg}}\simeq0.1942\,L$,  the  glass  state  fractures
further into four  ``frozen'' glass states of the  kind illustrated in
Fig.~\ref{critical}~(b), in  which one disk is confined  near a corner
of the  box.  Above $r=r_g\simeq0.1964\,L$, the system  can exist only
in the crystalline state.

\begin{figure}
\begin{center}
\includegraphics[width = 2.5in]{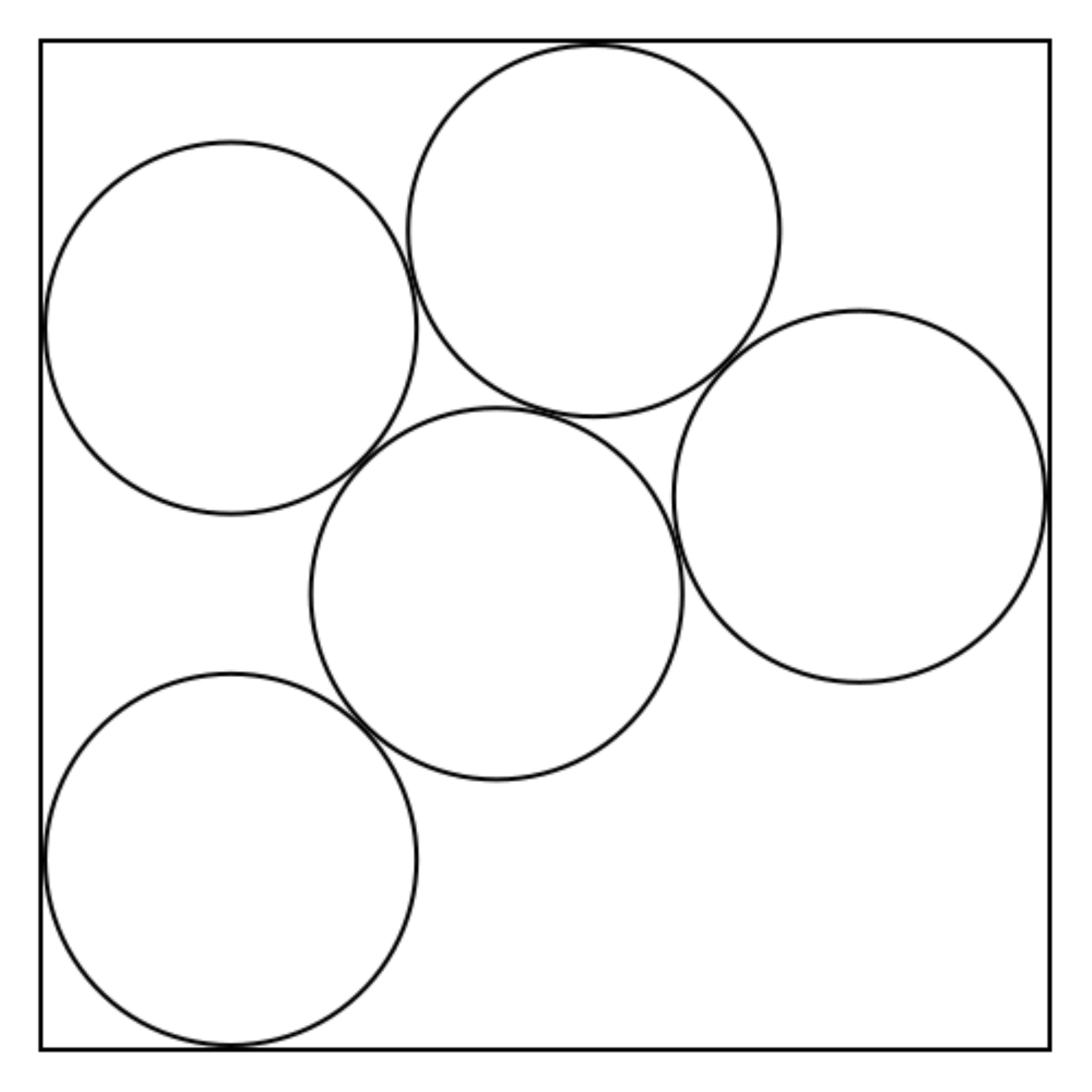}
\caption{The dead-end  configuration at  $r \simeq 0.1871  \,L$, which
  has   been   incorrectly   identified   as   a   saddle   point   in
  Refs.~\cite{Carlsson, Hinow}.}
\label{culdesac}
\end{center}
\end{figure}

It   may  be   noticed  that   $r_{\text{cg}}\simeq0.1863\,L$  differs
significantly   from  the   value   0.1871  stated   by  Carlsson   et
al.~\cite{Carlsson}.   We have  been  unable to  find  a path  between
glass-like  and crystal-like  metastable  states that  passes via  the
configuration proposed in their  paper.  We find, moreover, that their
proposed state  with $r\simeq0.1871\,L$ is  not a stationary  point of
the   softened   potential   energy   function   $E$   introduced   in
Ref.~\cite{Carlsson}:  instead,  it  is  a  minimum  of  $\lvert\nabla
E\rvert^2$ at  which $\nabla E\ne0$.  It is  a dead-end configuration,
illustrated in Fig.~\ref{culdesac}: it can be reached from the crystal
by the  steps in the  first two panels of  Fig.~\ref{critical}(a), but
progress to the glass state of  the third panel is not possible as the
central disk  cannot escape to the  edge of the square.   On the other
hand,  we can show  that our  own configuration  at $r\simeq0.1863\,L$
lies on  a path  between crystal-like and  glass-like states  and also
that  this configuration  corresponds  very precisely  to an  ordinary
saddle  point of~$E$.   Such  a  reaction path  is  illustrated by  an
animation provided in the supplement to this paper~\cite{animation}.

\section{Dynamics of the five-disk system}
\label{5diskdynamics}
As  for the  case of  two  disks, an  event-driven molecular  dynamics
algorithm \cite{Lubachevskyetnil}  was used to simulate  the motion of
the five-disk  system and calculate  the mean time of  passage between
metastable  states.  The initial  velocities of  the disks  were drawn
from the Maxwell-Boltzmann distribution.

A  very  simple  method was  used  in  our  work  to identify  when  a
transition had taken place.  For  the transition from crystal to glass
states,   the  simulation   is  started   in  a   typical  ``crystal''
configuration with one  disk [Fig.~\ref{critical}~(a), shaded] near to
the center  of the  box.  The shaded  disk's first collision  with any
wall is an unambiguous sign that the transition to the glass state has
occurred.   Transitions   between  metastable  glass   states  can  be
identified in a similar way.  From Fig.~\ref{critical}~(b), we can see
that a  transition has occurred  if a disk  [e.g.  the shaded  disk in
  Fig.~\ref{critical}~(b)] makes  a collision  with a wall  other than
the one it  was close to in the initial  configuration.  For each kind
of transition, the time of first occurrence of the diagnostic event is
recorded and the simulation restarted with random initial velocities.

Transition  state theory  requires us  to evaluate  $Z^{\ddagger}$ and
$Z$.   We shall  use the  procedure  introduced by  Salsburg and  Wood
\cite{SW} to  determine these as  it becomes essentially exact  as the
density approaches its maximum value (called $\phi_J$) appropriate for
a given inherent  state. Thus for the crystal  state $r_c=0.2071 \,L$,
so $\phi_J=5 \pi r_c^2/L^2 \approx 0.6738$. Let $l=(V/N)^{1/d}$ denote
the  average  spacing between  the  centers  of  the particles,  where
$V=L^d$.   The  Salsburg--Wood  approximation  is that  as  $\phi  \to
\phi_J$, $Z \sim l^{Nd}(1-\phi/\phi_J)^{Nd}$, where here $N=5$, $d=2$.
Similarly, as $\phi  \to \phi_{\text{Neck}}$ from below, $Z^{\ddagger}
\sim         l^{Nd-1}(1-\phi/\phi_{\text{Neck}})^{Nd-1}$         where
$\phi_{\text{Neck}}=0.5453$.   Hence,  according  to transition  state
theory, the  transition rate $R$ from  the crystal to  the glass state
should vary as
\begin{equation}
R=1/\tau \sim \frac{v}{l}\frac{(1-\phi/\phi_{\text{Neck}})^{N d-1}} {(1-\phi/\phi_J)^{N d}}.
\label{TSprediction}
\end{equation}
If  the reaction  coordinate is  fixed at  the value  it takes  in the
transition  state, the  hard-disk constraints  define  a configuration
space with  nine spatial dimensions  in our two-dimensional  hard disk
system.   Accordingly, the  constrained partition  function  should be
expected   to  vary   as  $(1-\phi/\phi_{\text{Neck}})^9$   using  the
procedure of Salsburg and Wood \cite{SW}.

\begin{figure}
\begin{center}
\includegraphics[width = 3.5in]{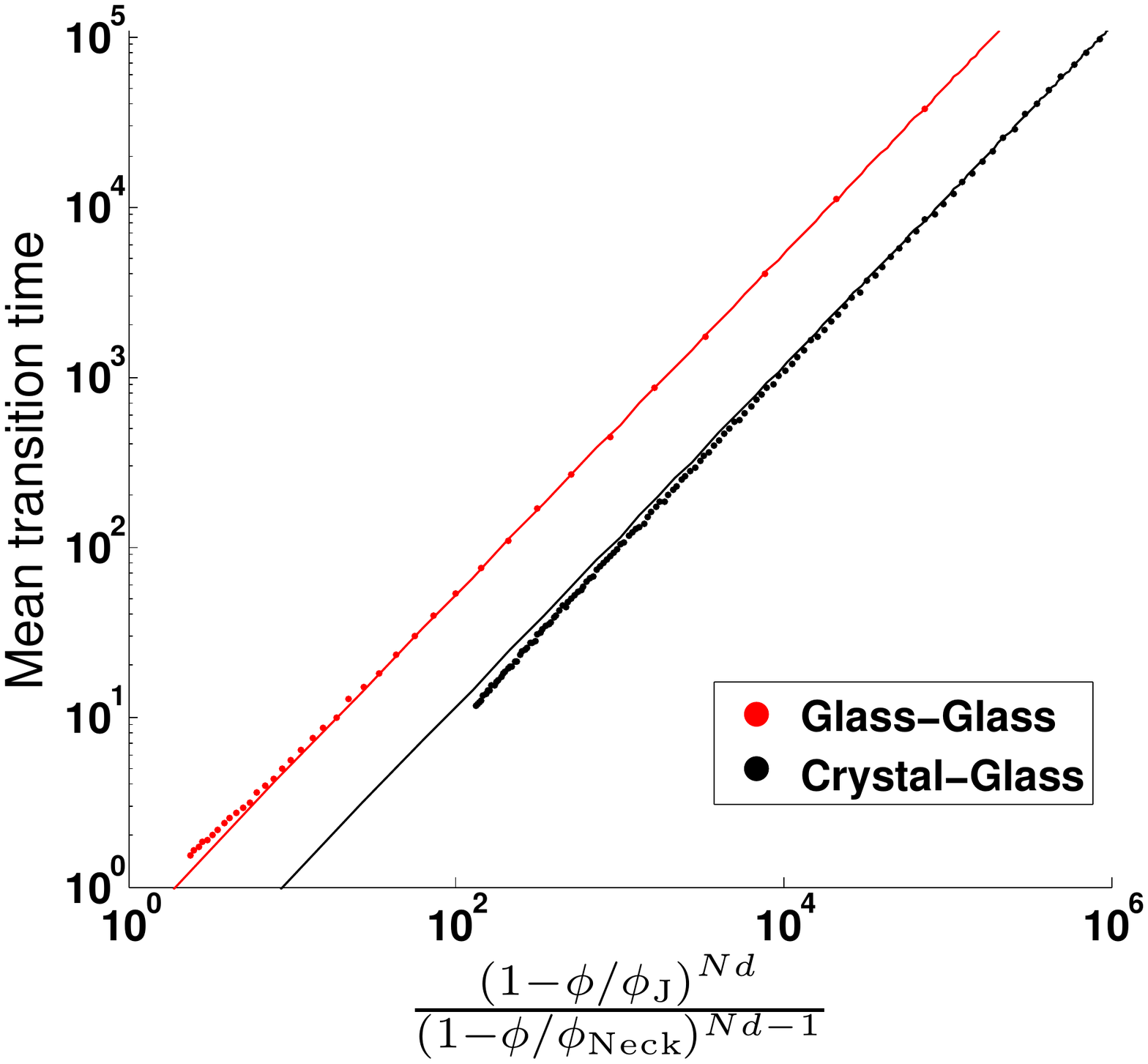}
\caption{(Color online)  Transition times between  pairs of metastable
  states.    Results from  molecular   dynamics are compared to  the
  predictions of transition state  theory for transitions between: (1)
  two glass  states just below the  ``glass--glass'' transition, where
  $\phi_J=  0.6061$  and   $\phi_{\text{Neck}}=0.5925$;  and  (2)  the
  crystalline   and   glass    states,   where   $\phi_J=0.6738$   and
  $\phi_{\text{Neck}}=0.5453$.   In each case,  $N=5$ and  $d=2$.  The
  error bars  for the molecular  dynamics results are  comparable with
  the size of the data points.}
\label{tau5}
\end{center}
\end{figure}

Our simulations to  test this for both the  glass glass transition and
the  crystal glass  transition  are plotted  in Fig.~\ref{tau5}.   The
agreement is excellent as $\phi \to \phi_{\text{Neck}}$.

\begin{figure}
\begin{center}
\includegraphics[width = 3.5in]{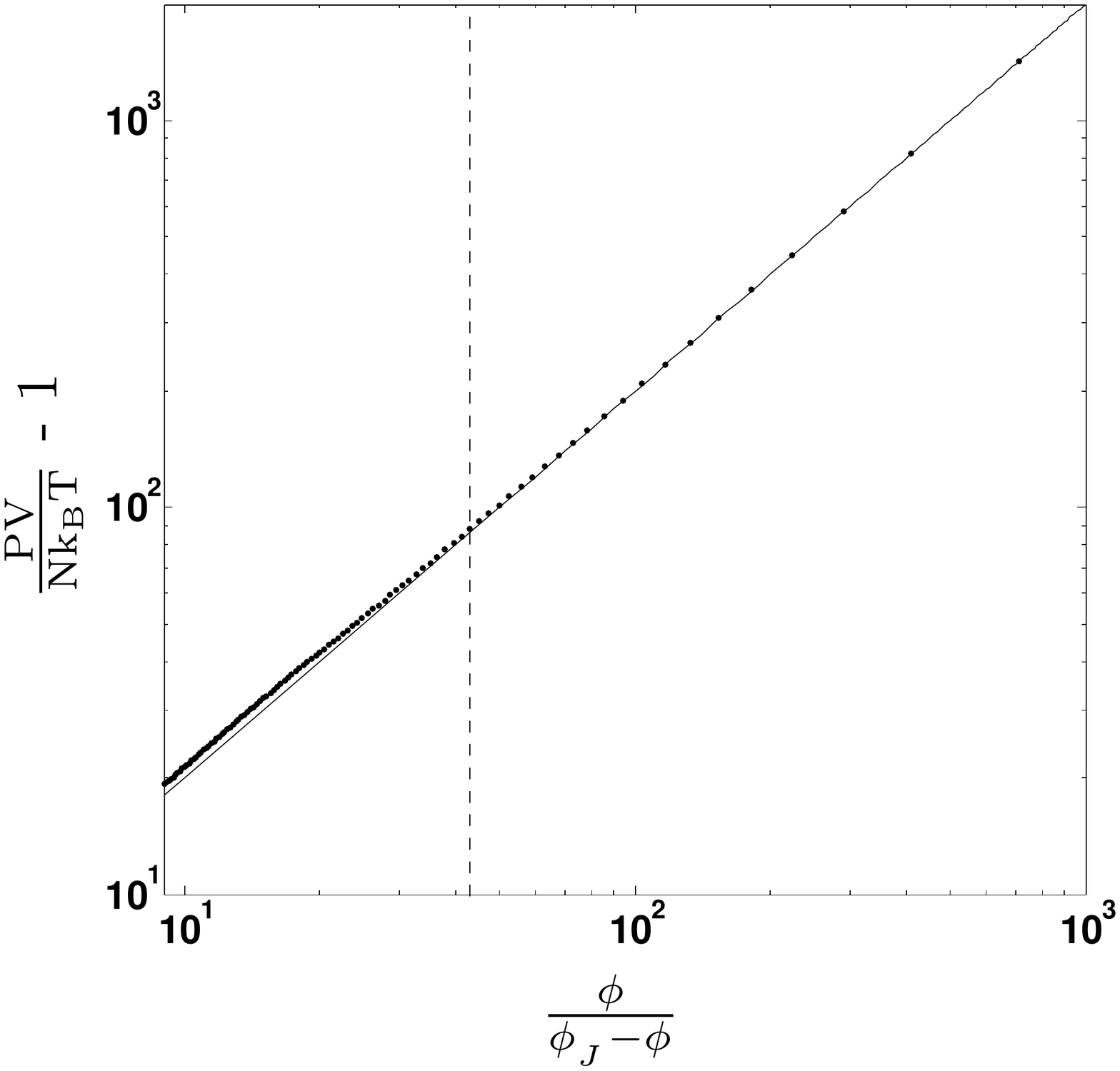}
\caption{Pressure  in the  glass  states. For  the  glass states,  the
  maximum  possible  packing fraction  is  at  $\phi_J= 0.6061$.   The
  packing fraction  at the neck  $\phi_{\text{Neck}}=0.5925$ separates
  the two kinds  of glassy state and is marked  by the dashed vertical
  line.   The glass  phase  only  exists down  to  a packing  fraction
  $0.5432$ as a  stable state and only data  obtained above this value
  are included.  The straight line is Eq.~(\ref{pressureSW}).}
\label{pressure}
\end{center}

\end{figure}

To further  examine the accuracy  of the Salsburg--Wood  procedure for
calculating  $Z$  and  $Z^{\ddagger}$,  we have  determined  from  our
molecular dynamics simulations the pressure of the system in the glass
states. The temperature was  obtained from the average kinetic energy,
using  $k_BT=m  \langle  v^2\rangle/2$.   The  results  are  shown  in
Fig.~\ref{pressure}.
The Salsburg--Wood approximation for $Z$ predicts that the pressure
\begin{equation}
\frac{PV}{Nk_BT}=1+d \frac{\phi}{\phi_J-\phi},
\label{pressureSW}
\end{equation}
on  using  the  relation  $P=k_BT \,\partial\ln  Z/\partial  V$.   The
straight line in Fig.~\ref{pressure}  represents the prediction of the
Salsburg--Wood     calculation      for     the     pressure,     i.e.
Eq.~(\ref{pressureSW}), and  is in perfect agreement with  the data as
$\phi \to \phi_J$.   Notice that at the neck,  $\phi=0.5925$, which is
indicated  by the  vertical  dashed line  in Fig.~\ref{pressure},  the
singularities  are so  mild  as to  be  invisible, which  means it  is
adequate to use in Eq.~(\ref{TSprediction}) the form of $Z$ valid near
$\phi_J$, even  for $\phi$ close to  $\phi_{\text{Neck}}$.  (Also, the
expression  for $\tau$  is dominated  by the  form  of $Z^{\ddagger}$,
which is  rapidly approaching  zero as $\phi  \to \phi_{\text{Neck}}$,
while $Z$ is there only slowly varying.)

\section{Large numbers of spheres or disks}
\label{extensionlargeN}
In  our studies  of two  and  five disks  we found  that a  transition
between states in a region containing $N$ particles generally requires
coordinated  motion of  all  the  $N$ particles  in  order to  squeeze
through the neck in the phase space. The rate at which this will occur
was     given     by    the     transition     state    formula     of
Eq.~(\ref{TSprediction}). In this section, we examine the consequences
of assuming that  the formula can be extended  to systems containing a
large number $N$ of spheres or disks.

We shall first suppose that one is at a packing fraction below that of
the  neck out  of an  inherent  state whose  largest density  is at  a
packing fraction $\phi_J$ and that $\phi_{\text{Neck}}$ is the highest
packing fraction below  $\phi_J$ at which a neck  first opens to allow
escape from  the inherent  state, and that  one is in  a configuration
close to that of the inherent state.  Furthermore we shall assume that
when $N$ is large,
\begin{equation}
\phi_J-\phi_{\text{Neck}}=a \phi_J/N,
\label{VFassumption}
\end{equation}
where  $a$ is  a  positive  constant of  O(1).  The assumption  behind
Eq.~(\ref{VFassumption})  is  that escape  from  a  jammed state  will
become possible if the volume of  the system is increased by an amount
of the order of the volume  of a single sphere.  With this assumption,
and taking $N$ to be large, Eq. (\ref{TSprediction}) reduces to
\begin{equation}
\frac{1}{\tau}=\frac{1}{\tau_0}\left[1- \frac {a \phi}{(1-\phi/\phi_J)\phi_J N}\right]^{Nd-1},
\end{equation}
where  $\tau_0=l(1-\phi/\phi_J)/v$ denotes  the  typical time  between
collisions of the disks.  Then as $N \to \infty$,
\begin{equation}
\tau=\tau_0 \exp \left[\frac{a d \phi}{(\phi_J-\phi)}\right],
\label{VFT}
\end{equation}
which is the Vogel--Fulcher--Tammann formula.

Given a particular  configuration of the $N$ particles  with a packing
fraction $\phi$ we  need to know the packing  fraction $\phi_J$ of the
nearby inherent  state close to  the initial configuration.   In other
words,  we need  the  Stillinger  map to  the  jammed inherent  states
\cite{Stillingermap,Bowles}. For the problem of disks moving in a long
narrow channel such a  map was explicitly constructed in \cite{Bowles}
and  the function  $\phi_J(\phi)$ exhibited.   Except for  quite small
values of $\phi$, $\phi_J(\phi)$ is essentially a constant independent
of $\phi$  and close to the  largest packing fraction  possible in the
system. The  map is  similar to  what would have  been obtained  in an
extremely  rapid  compression. The  relaxation  times  $\tau$ in  this
narrow channel system are consistent with Eq.~(\ref{VFT}) \cite{Ivan}.
It has proved  possible to identify the inherent  states and the necks
which have to  be squeezed through to escape from  the vicinity of the
inherent states and as a  consequence the value of the coefficient $a$
can be explicitly determined for this system \cite{GodfreyMoore}.

In dimensions  $d >1$ much less  can be said with  certainty.  Fits of
the alpha  relaxation time  to the VFT  formula for  three dimensional
hard spheres were made by Brambilla et al.  \cite{Brambilla} and a fit
was achieved with  a value of $\phi_J \approx  0.615$.  One might have
expected that the appropriate value of $\phi_J$ if the map from $\phi$
to the inherent state is essentially a rapid compression would be that
of random close packing, $\phi_{\text{rcp}} \approx 0.64$.  The result
that  $\phi_J \approx  0.615$ was  obtained for  studies of  $\tau$ at
$\phi \le 0.6$ and it might require data at larger values of $\phi$ to
produce $\phi_J$ values closer to $\phi_{\text{rcp}}$.

We  have  been assuming  that  the Stillinger  map  in  two and  three
dimensions, $\phi_J(\phi)$,  is essentially a  constant independent of
$\phi$.  This  lack of any  $\phi$ dependence of  $\phi_J(\phi)$ seems
unlikely according to the studies in \cite{Ozawa, Chaudhuri}.  Suppose
that instead the Stillinger map  in two and three dimensions takes the
form, for $\phi$ close to $\phi_{\text{rcp}}$,
\begin{equation}
\phi_J(\phi) \approx \phi +B(\phi_{\text{rcp}}-\phi)^{\delta},
\label{phiJphi}
\end{equation}
with $B>0$.   To test  this supposition, one  would need to  start the
rapid compression from the well-equilibrated fluid system at a packing
fraction near $\phi_{\text{rcp}}$.  Producing this initial state would
be difficult.  If Eq.~(\ref{phiJphi}) is  valid, it would lead  to the
following expression for the alpha relaxation time
\begin{equation}
\tau_{\alpha}(\phi)=\tau_0 \exp\left[\frac{A}{(\phi_{\text{rcp}}-\phi)^{\delta}}\right].
\end{equation}
In Ref. \cite{Brambilla}, a good fit was obtained with $\delta =2$ and
a value for $\phi_{\text{rcp}}\approx 0.64$ --- a commonly quoted value.

In  words, Eq.~(\ref{phiJphi})  states  that if  one  starts from  the
equilibrated   system  at   a   packing  fraction   $\phi$  close   to
$\phi_{\text{rcp}}$,  then the  rapid compression  (or  the Stillinger
map) finds a jammed state whose packing fraction $\phi_J$ only differs
from   $\phi$   by    a   quantity   of   order   $(\phi_{\text{rcp}}-
\phi)^{\delta}$,  which  is  small  when $\delta  >1$.   The  physical
implication is  that equilibrated systems  at such high  densities are
always  close to  a jammed  state.  However,  Eq.~(\ref{phiJphi}) also
assumes that $\phi_{\text{rcp}}$ is a well-defined density and this is
contentious  \cite{Torquato}.  Notice that  our difficulties  in using
Eq.~(\ref{VFT}) stem from just not  knowing the form of the Stillinger
map  $\phi_J(\phi)$  for two  and  three  dimensional  systems. It  is
possible that it takes a  form that would leave $\tau_{\alpha}$ finite
for  all  $\phi$ less  than  that of  the  maximum  density.  In  this
situation it  could be that for $\phi$  well below $\phi_{\text{rcp}}$
$\tau_{\alpha}$   might  appear   to   be  diverging   as  $\phi   \to
\phi_{\text{rcp}}$,   but  if  studies   could  be   performed  nearer
$\phi_{\text{rcp}}$  the  relaxation  times  would be  very  long  but
finite.

In conclusion  we have  shown that the  long relaxation times  seen in
small systems  of two and five disks  confined in a square  are due to
squeezing through necks in  configuration space, and can be understood
quantitatively  with the  aid  of transition  state  theory.  We  have
suggested that a  similar mechanism might be relevant  to hard spheres
in  higher dimensions  and could  lead either  to the  VFT  formula or
possibly a generalization of it.

We should like  to thank Chris Fullerton for  many useful discussions,
and Steve Teitel for supplying useful references.

\end{document}